\documentclass[11pt,a4paper]{article}
\usepackage{amssymb,fullpage}
\usepackage[super,sort&compress]{natbib}
\usepackage{amsfonts,amsbsy}
\usepackage{graphicx}
\let\ssection=\section
\renewcommand{\section}{\setcounter{equation}{0}\ssection}

\def\e{\mathrm{e}}
\def\i{\mathrm{i}}
\def\d{\mathrm{d}}

\def\beq{\begin{equation}}
\def\eeq{\end{equation}}
\def\Re{\mathrm{Re}\,}

\def\eps{\epsilon}

\def\vzeta{\mbox{\boldmath{$\zeta$}}}
\def\sgn{\mathrm{sgn} \,}

\def\Bi{\mathrm{Bi}}
\def\Ai{\mathrm{Ai}}
\def\ph{\mathrm{ph}\,}
\def\sig{\varsigma}
\def\Res{\mathrm{Res}\,}
\def\a{a}
\def\b{b}
\def\invb{\mu}
\def\bx{\boldsymbol{x}}
\def\bk{\boldsymbol{k}}
\def\bU{\boldsymbol{U}}

\newcommand{\dt}[2]{\frac{\mathrm{d} #1}{\mathrm{d} #2}}
\newcommand{\eqn}[1]{(\ref{eqn:#1})}
\newcommand{\lab}[1]{\label{eqn:#1}}
\newcommand{\inter}[1]{\quad \textrm{#1} \quad}

\def\Xint#1{\mathchoice
{\XXint\displaystyle\textstyle{#1}}%
{\XXint\textstyle\scriptstyle{#1}}%
{\XXint\scriptstyle\scriptscriptstyle{#1}}%
{\XXint\scriptscriptstyle\scriptscriptstyle{#1}}%
\!\int}
\def\XXint#1#2#3{{\setbox0=\hbox{$#1{#2#3}{\int}$}
\vcenter{\hbox{$#2#3$}}\kern-.5\wd0}}

\def\dashint{\Xint-}

\title{Elliptical instability of a rapidly rotating, strongly stratified fluid}
\author{J M Aspden and J Vanneste}
\date{School of Mathematics and Maxwell Institute for Mathematical Sciences \\
University of Edinburgh, Edinburgh EH9 3JZ, UK}

\begin{document}
\maketitle


\noindent
The elliptical instability of a rotating stratified fluid is examined in the regime of small Rossby number and order-one Burger number corresponding to rapid rotation and strong stratification. The Floquet problem describing the linear growth of disturbances to an unbounded, uniform-vorticity elliptical flow is solved using exponential asymptotics. The results demonstrate that the flow is unstable for arbitrarily strong rotation and stratification; in particular, both cyclonic and anticyclonic flows are unstable. The instability is weak, however, with growth rates that are exponentially small in the Rossby number. The analytic expression obtained for the growth rate elucidates its dependence on the Burger number and on the eccentricity of the elliptical flow. It explains in particular the weakness of the instability of cyclonic flows, with growth rates that are only a small fraction of those obtained for the corresponding anticyclonic flows. The asymptotic results are confirmed by numerical solutions of Floquet problem.

\section{Introduction}
Elliptical instability, the three-dimensional instability of two-dimensional flows with elliptical streamlines, has been the focus of a great deal of research activity. The review by \citet{kers02} discusses the main results up to 2002 and emphasises the relevance of elliptical instability to a broad range of applications. One of these is the instability of two-dimensional vortices that are deformed elliptically by a large-scale strain flow.  This is especially important for the dynamics of the atmosphere and ocean, since this is characterised by an abundance of vortices that are deformed through either mutual interactions or the effect of large-scale flows. In this context, however, the planetary rotation and density stratification need to be taken into account. 

Rotation and stratification clearly exert a strong influence on elliptical instability: since this stems from the parametric resonance between the periodic fluctuations associated with the elliptical motion and the free waves supported by the flow, the dispersion relation of these waves is critical. In the presence of rotation and stratification, the waves are inertia-gravity waves whose frequency is bounded from below by the minimum of the Coriolis parameter $f$ and Brunt--V\"ais\"al\"a frequency $N$. As a consequence, a vortex of fixed vorticity ceases to be unstable by the subharmonic instability responsible for the simplest form of elliptical instability when $f$ and $N$ exceeds a certain threshold. As these parameters increase further, instabilities are limited to resonances of higher and higher order, leading to decreasing growth rates. This was clearly demonstrated by \citet{miya93} on the basis of numerical solutions of the Floquet problem that models elliptic instability. Further numerical results were obtained by \citet{mcwi-yavn} who concentrated on the regime of rapid rotation and strong stratification with $N>f$ most relevant to the atmosphere and ocean. Their broad motivation was the role that instabilities play in the generation of inertia-gravity-wave-like motion, and the resulting breakdown of the nearly geostrophic and hydrostatic balance that is typical of much of the atmosphere and ocean. The present paper shares the same motivation. It re-examines the elliptical instability of a rotating stratified fluid and derives explicit analytical results in the limit of fast rotation $f \gg \Omega$ and strong stratification $N \gg \Omega$, where $\Omega$ denotes the (relative) vorticity of the flow. 

Several recent papers\citep{mole-et-al05,v-yavn07,mcwi-et-al04,plou-et-al} have demonstrated in specific examples that instabilities of well-balanced basic flows to inertia-gravity-wave perturbations (or perturbations related to similar fast waves) have growth rates that are exponentially small in the Rossby number, here proportional to $\Omega / f \ll 1$.\footnote{Note that we use the convential form of Rossby number rather than its inverse as used in Ref.\ {\citep{miya93}}.} Theoretical arguments\cite{v-yavn04,v08} indicate that this is a generic property, and the elliptical instability examined in this paper is no exception. In this case, the exponential smallness can be roughly understood by noting that, in the manner typical of parametric instabilities\citep{bend-orsz}, the growth rates of the elliptical instability can be expected to be proportional to $\Omega^n$, where $n$ is the order of the resonance. Since, as pointed out by \citet{miya93}, the minimum $n$ is of the order $f/\Omega$ (for $N > f$), this leads to the conclusion that growth rates are  beyond all orders in the Rossby number. To go further than this rough argument and provide an estimate for the growth rate requires the exponential-asymptotics analysis of the Floquet problem relevant to the elliptical instability. We carry out this analysis and, rather than relying on general asymptotic results for Hill's equations\citep{wein-kell87}, directly relate the growth of solutions to the occurrence of a Stokes phenomenon\citep{ablo-foka} which we capture using a combination of WKB expansion and matched asymptotics in complex time. The analytical results are confirmed by the numerical solutions of the Floquet problem. 

One of the issues which our treatment clarifies is the difference between the instability of cyclonic and anticyclonic vortices. Cyclones have been recognised as less unstable than anticyclones, to the extent that \citet{mcwi-yavn} considered only the instability of the latter. We show that the cyclones are in fact linearly unstable, with growth rates that have the same exponential dependence on the Rossby number as the corresponding anticyclones but differ by a factor which, although formally of order one, turns out to be numerically very small.

The plan of this paper is as follows. In section 2, we formulate the problem of elliptical instability in a rotating stratified fluid modelled using the Boussinesq approximation. We use the simplest instance of elliptical instability, that of an unbounded elliptical vortex with uniform vorticity. This makes it possible to seek global solutions in the form of plane waves with time-periodic wavevector and an amplitude that satisfies a Hill's equation. (Results for this particular case have a much broader appeal, however, since an identical Hill's equation arises when the stability of more general elliptical vortices is examined locally using the geometric-optics technique.\citep{frie-lipt,ledi00}) The Floquet problem associated with the Hill's equation is solved asymptotically in section 3 in the limit of fast rotation and strong stratification, with the  eccentricity of the elliptical streamlines assumed of order one. For simplicity, we also make the hydrostatic approximation assuming that $N \gg f$ and an order-one Burger number. The asymptotic derivation is only sketched in section 3, with technical details relegated to Appendix A.  The asymptotic results are confirmed by direct numerical solution of the Floquet problem in section 4. The effect of a finite $N/f$ is also briefly examined there.

\section{Formulation}

We consider the stability of a horizontal elliptical flow in a three-dimensional stratified fluid, with constant Brunt--V\"ais\"al\"a frequency $N$, rotating about the vertical axis at rate $f/2 > 0$. The flow's streamfunction, velocity and vorticity are written as
\beq \lab{flow}
\Psi = - \frac{1}{2} \left(\b x^2 + \a y^2\right), \quad  \bU=(\a y,-\b x,0) \inter{and} \Omega=a+b,
\eeq
where $ab > 0$. We define
\[
\sig = \sgn \a = \sgn \b
\]
and note that the flow is anticyclonic for $\sig = 1$, and cyclonic for $\sig=-1$. Three dimensionless parameters characterise the flow, namely
\beq \lab{param}
e=\sqrt{\a/\b}, \quad \eps=\sqrt{\a \b}/f \inter{and} f/N,
\eeq
which are recognised as the aspect ratio of the elliptical flow,  a Rossby number and  the Prandtl ratio. We assume that $e > 1$ without loss of generality.

Perturbations to the flow \eqn{flow} take the form of plane waves with time-dependent wavevector, with each field written as
\[
u(\bx,t) = \hat{u}(t) \e^{\i \bk (t) \cdot \bx},
\]
where the wavevector $\bk=(k,l,m)$ satisfies
\beq \lab{vectorevol}
\dot k = b l, \quad \dot l = - a k \inter{and} \dot m = 0,
\eeq
the overdot denoting differentiation with respect to $t$. 
In what follows, we use a dimensionless time variable obtained by taking $(\a \b)^{-1/2}$ as a reference time. In terms of this variable, the solutions to \eqn{vectorevol} have the simple form
\beq \lab{wavevec}
k = k_0 \cos t, \quad l = - \sig  e k_0 \sin t \inter{and} m=m_0,
\eeq
where $k_0$ and $m_0$ are constant. The stability of \eqn{flow} depends on the behaviour of the amplitudes $\hat{u}(t)$, $\hat{v}(t)$, etc.\ as $t \to \infty$.
These satisfy a set of ordinary differential equations with time-periodic coefficients. Following \citet{mcwi-yavn}, this set can be conveniently reduced to a single second-order equation for the amplitude of the vertical component of the vorticity $\hat{\zeta}=\i (l \hat{v} -k \hat{u})$. Assuming that the perturbation potential vorticity vanishes, this equation reduces to
\beq \lab{nonhydro}
\ddot \zeta + \frac{2 \sig kl m^2(e-e^{-1})}{\kappa^2(k^2+l^2)} \dot \zeta + \frac{1}{\eps^2} \left[ \left(1-\sig \eps (e +e^{-1}\right) \left( 1-\frac{2 \sig \eps e k_0^2}{k^2+l^2}\right) \frac{m^2}{\kappa^2} + \frac{N^2 (k^2 + l^2)}{f^2 \kappa^2} \right] \zeta = 0,
\eeq
where $\kappa^2 = k^2 + l^2 + m^2$ and we have omitted the hat on the amplitude $\zeta$. Four dimensionless parameters appear in this equation: the three flow parameters \eqn{param}, and the initial aspect ratio $m_0/k_0$ of the perturbation. Note that anticyclonic flows (with $\varsigma =1)$ are susceptible to centrifugal instability (e.g.\ Ref.\citep{kloo-et-al}) when the relative vorticity exceeds $f$, that is, for $\eps (e+e^{-1}) > 1$. Since we focus on the regime $\eps \ll 1$ we do not this consider this instability further. 

Most of this paper  focuses on a limiting case of \eqn{nonhydro} obtained by making the hydrostatic approximation. This assumes that $m_0 \gg k_0 $ and $N \gg f$ while
\beq \lab{invb}
\invb = \frac{f m_0}{N k_0}=O(1).
\eeq
This is the regime most relevant to the dynamics of the atmosphere and oceans since the condition $N \gg f$ is verified while, as we demonstrate below, the largest growth rates of the elliptical instability correspond to $\invb=O(1)$. The parameter $\mu$ can be recognised as the inverse square root of a Burger number; it can be interpreted as the aspect ratio of the perturbation scaled by $f/N$ as is natural in rapidly rotating, strongly stratified fluids.

In the hydrostatic approximation, $\kappa^2$ is approximated by $m^2$, and
\eqn{nonhydro} reduces to
\beq \lab{zeta1}
\ddot \zeta + \frac{2 \sig kl(e-e^{-1})}{k^2+l^2} \dot \zeta + \frac{1}{\eps^2} \left[ \left(1-\sig \eps (e +e^{-1}\right) \left( 1-\frac{2 \sig \eps e k_0^2}{k^2+l^2}\right) + \frac{N^2 (k^2 + l^2)}{f^2 m^2} \right] \zeta = 0.
\eeq
Using \eqn{wavevec} and defining $\psi>0$ by
\beq \lab{psi}
e^2 = 1+\psi^2,
\eeq
we rewrite this equation as
\beq \lab{zeta2}
\ddot \zeta -  p(t) \dot \zeta +
\frac{1}{\eps^2} \left[\omega^2(t) - \eps q(t) + \eps^2 r(t) \right] \zeta = 0.
\eeq
Here
\beq \lab{omega}
\omega^2 = 1 + \frac{N^2(k^2 + l^2)}{f^2 m^2} = 1 + \invb^{-2} (1+\psi^2 \sin^2 t),
\eeq
can be recognised as the square of the inertia-gravity-wave frequency (non-dimensionalised by $f$). We have also introduced
\beq \lab{fg}
p(t)= \frac{\psi^2 \sin(2 t)}{1+\psi^2 \sin^2 t}, \quad
q(t) = \sig \left(e + e^{-1} + \frac{2 e}{1+\psi^2 \sin^2 t}\right) \inter{and}
r(t) = \frac{2(e^2+1)}{1+\psi^2 \sin^2 t}.
\eeq

Equation \eqn{zeta2} is a Hill equation, with coefficients that are $\pi$-periodic in $t$. Its stability is determined using the Floquet theory for Hill equations\citep{bend-orsz}. Briefly, if
$\vzeta(t) = (\zeta_1(t),\zeta_2(t))^\mathrm{T}$ is a column vector of independent solutions,
\[
\vzeta(t+\pi) = M \vzeta(t)
\]
for some constant matrix $M$. The eigenvalues $\lambda$ of $M$ are then the Floquet multipliers, and two fundamental solutions can be found for which
\beq
\zeta(t) =\e^{\sigma t} \phi(t),
\eeq
where
\beq \lab{siglam}
\sigma = \frac{1}{\pi} \log \lambda
\eeq
is the Floquet exponent, and $\phi(t)$ is $\pi$-periodic. Note that the form of the coefficient of $\dot \zeta$ ensures that the two multipliers satisfy $\lambda_1 \lambda_2 = 1$. Instability occurs when one of these is such that $|\lambda|>1$ or, equivalently, $\Re \sigma > 0$. The matrix $M$ is computed by relating $\vzeta$ and $\dot {\vzeta}$ at two times $t$ and $t+\pi$. Here we choose $t = -\pi/2$ and compute $M$ as
\beq \lab{m1}
M = [ \vzeta(\pi/2),\dot{\vzeta}(\pi/2)] [ \vzeta(-\pi/2),\dot{\vzeta}(-\pi/2)]^{-1}.
\eeq

Our aim is to determine the largest values of the growth rate $\Re \sigma$ for \eqn{zeta2} analytically in the fast-rotation regime $\eps \ll 1$, with $N \gg f$, $\invb=O(1)$ and $\psi=O(1)$. In this regime \eqn{zeta2} ressembles the Hill  equations with large parameters whose stability has been studied by  \citet{wein-kell87} using a mapping to parabolic cylinder functions.
However, there are difficulties in applying their results directly, related to the presence of a first derivative term that is singular for the complex values of $t$ such that $k^2+l^2=0$. We have therefore found it simpler to develop a different approach, combining WKB analysis with complex-time matching.
This approach, which has the advantage of demonstrating the link between the instability and the Stokes phenomenon\citep{ablo-foka,pari-wood}, is described in the next section and in Appendix \ref{sec:expasy}. The analytic results obtained there are confirmed and extended to finite $N/f$ in section 4 by solving \eqn{zeta2} numerically.

\section{WKB analysis}

A WKB solution of the form
\beq \lab{wkb1}
\zeta = A(t) \e^{\i \theta(t)/\eps}
\eeq
can be introduced into \eqn{zeta2}, and the (real) functions $A(t)$ and $\theta(t)$ can be derived by expansion in powers of $\eps$.
At leading order, we find that
\beq \lab{theta0}
\theta_0(t) = \int_{-\pi/2}^t \omega(t') \, \d t'.
\eeq
At the next order, we have
\begin{eqnarray}
\frac{\dot A_0}{A_0} &=& - \frac{\dot \omega}{2 \omega} + \frac{p}{2}, \lab{a}
 \\
\dot \theta_1 &=& - \frac{q}{2 \omega}. \lab{theta1}
\end{eqnarray}
We note that $A_0(t)$ is $\pi$-periodic; this is also true for higher-order corrections, so that there is no instability to any algebraic order in $\eps$: the fundamental solutions are given by $\zeta$ in \eqn{wkb1} and its complex conjugate, and the Floquet multipliers are simply $\lambda = \pm \exp(\i \theta(\pi/2))$ (taking $\theta(-\pi/2)=0$). Instability is necessarily an exponentially small effect; furthermore, it can only occur for values of the parameters such that $\lambda$ is exponentially close to $\pm 1$. This is because the condition for instability $| \lambda | > 1$ requires the two multipliers to be purely real; however, they are complex conjugate to all orders in $\eps$, a property which persists in the presence of a small perturbation in the non-degenerate cases $\lambda \not= \pm 1$.

Computations detailed below show that the solution defined by \eqn{wkb1} for $-\pi/2 \le t < -\delta$, with $\delta \ll 1$, switches on an exponentially small term as the Stokes line $\Re t =0$ is crossed. Denoting by $S$ the corresponding Stokes multiplier, which is exponentially small in $\eps$, this implies that the pair of solution
\beq \lab{wkb-}
\zeta = A(t) \e^{\i \theta(t)/\eps} \inter{and} \bar \zeta = A(t) \e^{-\i \theta(t)/\eps}
\eeq
valid for $-\pi/2 \le t < -\delta$ (for some $\eps^{1/2} \ll \delta \ll 1$) becomes
\beq \lab{wkb+}
\zeta = A(t) \left[ \e^{\i \theta(t)/\eps}  + S \e^{-\i \theta(t)/\eps} \right]  \inter{and} \bar \zeta = A(t)  \left[ \e^{-\i \theta(t)/\eps}  + \bar S \e^{\i \theta(t)/\eps} \right]
\eeq
for $\delta < t \le \pi/2$. Taking $\vzeta=(\zeta,\bar \zeta)^\mathrm{T}$, we compute the matrix $M$ in \eqn{m1}. We first have that
\[
[\vzeta(\pi/2),\dot{\vzeta}(\pi/2)] = \left(
\begin{array}{cc}
A (\e^{\i \theta/\eps} + S  \e^{-\i \theta/\eps}) & (\i \eps^{-1} \dot \theta A + \dot A)  \e^{\i \theta/\eps} + S (-\i \eps^{-1} \dot \theta  A+ \dot A) \e^{-\i \theta/\eps} \\
A (\e^{-\i \theta/\eps} + \bar S  \e^{\i \theta/\eps})& (-\i \eps^{-1} \dot \theta A + \dot A)  \e^{-\i \theta/\eps} + \bar S (\i \eps^{-1} \dot \theta  A + \dot A)\e^{\i \theta/\eps}
\end{array}
\right),
\]
where $A$, $\dot A$, $\theta$ and $\dot \theta$ are evaluated at $t=\pi/2$. Similarly,
\[
[\vzeta(-\pi/2),\dot{\vzeta}(-\pi/2)] = \left(
\begin{array}{cc}
A & \i \eps^{-1} \dot \theta A + \dot A \\
A & -\i \eps^{-1} \dot \theta A + \dot A
\end{array}
\right).
\]
Here $A$, $\dot A$ and $\dot \theta$ can be evaluated at $t=\pi/2$, as above, since their values at $t = \pm \pi/2$ coincide. Computing \eqn{m1} gives the simple result
\beq \lab{m2}
M = \left(
\begin{array}{cc}
\e^{\i \theta/\eps} & S \e^{-\i \theta/\eps} \\
\bar S \e^{\i \theta/\eps} & \e^{-\i \theta/\eps}
\end{array}
\right).
\eeq
Here $\theta=\theta(\pi/2)$ or, more generally, $\theta=\theta(\pi/2)-\theta(-\pi/2)$ which accommodates any convention for the arbitrary choice of $\theta(-\pi/2)$.

\begin{center}
\begin{figure}[t]
\begin{tabular}{cc}
$\alpha$ & $\beta$ \\
\includegraphics[width=0.5\textwidth]{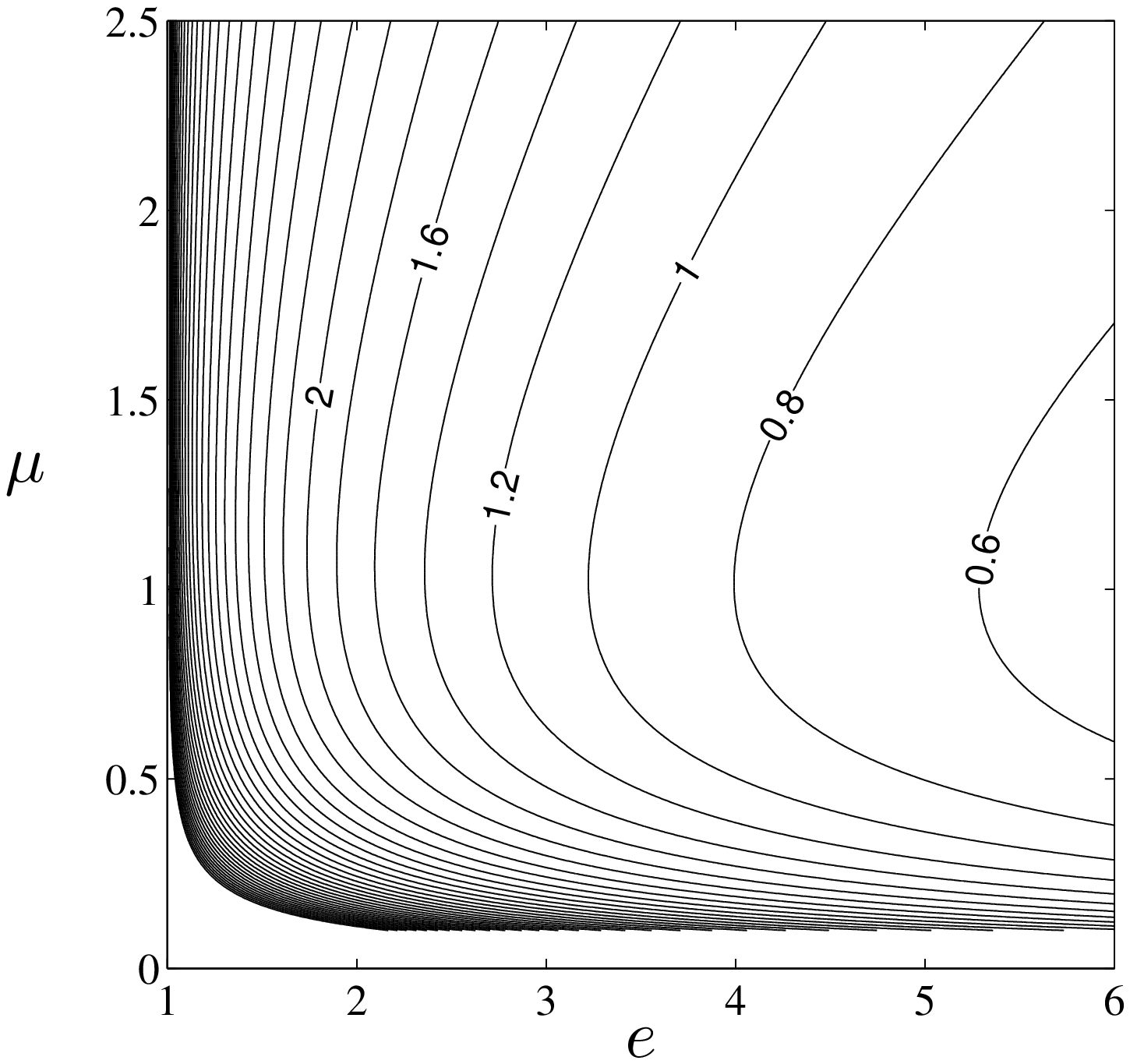}
&
\includegraphics[width=0.5\textwidth]{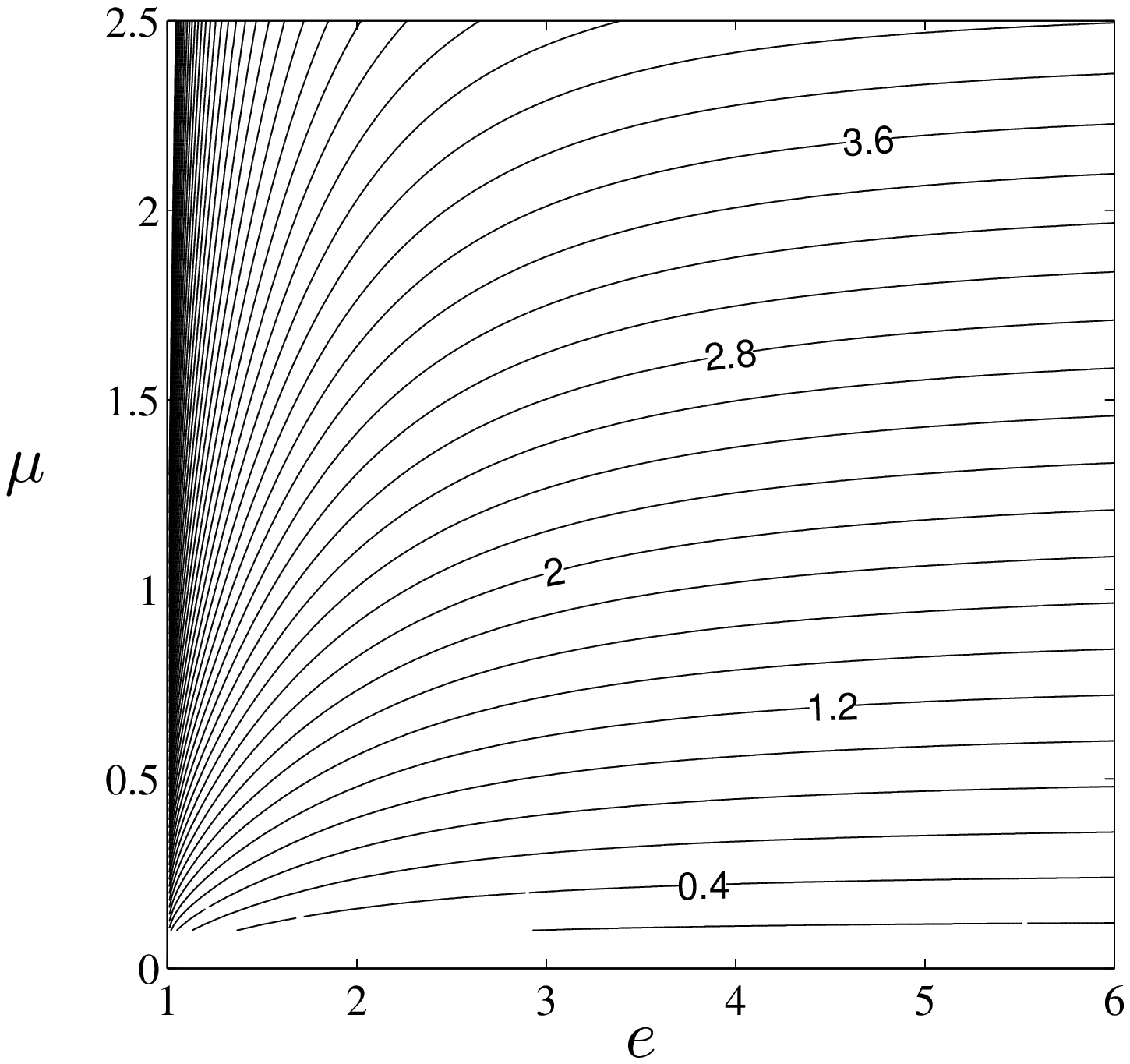}
\end{tabular}
\caption{Parameters $\alpha$ and $\beta$ governing the maximum growth rates according to \eqn{sigma}--\eqn{S} as functions of $e$ and $\mu$.}
\label{alpha_beta_contours}
\end{figure}
\end{center}

Suppose now that the parameters are such that
\beq \lab{eitheta}
\e^{\i \theta/\eps} = \pm (1+\i T) + O(T^2)
\eeq
for some $T \in \mathbb{R}$  of a similar order of magnitude as $S$. The Floquet multipliers obtained from \eqn{m1} are then given by
\beq \lab{lambda}
\lambda = \pm \left( 1 + \sqrt{|S|^2-T^2} \right) + O(|S|^2) \inter{and}
\lambda = \pm \left( 1 - \sqrt{|S|^2-T^2} \right) + O(|S|^2).
\eeq
Clearly, one of these multipliers satsifies $|\lambda|>1$, and the  flow is unstable, provided that $-|S| \le T \le |S|$, that is, in exponentially narrow instability bands. The corresponding growth rate $\sigma=\sqrt{|S|^2-T^2}/\pi + O(|S|^2)$ is maximum at the centre of these bands, for $T=0$, and is given by
\beq \lab{sigma}
\sigma_\mathrm{max} \sim  \frac{|S|}{\pi}.
\eeq
The computation of $S$ is carried out in Appendix \ref{sec:expasy}. There we show that
\beq \lab{S}
|S| = \e^{-\alpha/\eps+\sig \beta},
\eeq
where
\begin{eqnarray}
\alpha &=& \frac{2}{\invb} \int_0^{\sqrt{1+\invb^2}/\psi} \sqrt{\frac{1+\invb^2-\psi^2 x^2}{1+x^2}} \, \d x \quad \textrm{and} \lab{alpha} \\
\beta &=& \invb \, \dashint_0^{\sqrt{1+\invb^2}/\psi} \left( e + e^{-1} + \frac{2e}{1-\psi^2 x^2} \right) \frac{\d x}{\sqrt{(1+\invb^2-\psi^2 x^2)(1+x^2)}}. \lab{beta}
\end{eqnarray}
Here $\dashint$ denotes the Cauchy principal value of the integral, whose integrand is singular at $x=1/\psi$.

Figure \ref{alpha_beta_contours} shows the values of $\alpha$ and $\beta$ as functions of $e$ and $\invb$. Some conclusions can be drawn from the figure and the examination of the explicit expressions \eqn{alpha}--\eqn{beta}. First, $\alpha \to \infty$ in the limits of both small and large $\invb$; specifically  $\alpha = O(\mu^{-1}$) as $\invb \to 0$ and $\alpha = O(\log \mu)$ as $\invb \to \infty$. This suggests, as is confirmed by Figure \ref{alpha_beta_contours}, that the largest growth rates are attained for $\mu=O(1)$. Thus, the aspect ratio of the perturbations that grow as a result of the elliptical instability of vortices should be expected to be the Prandlt ratio: $m_0/k_0 = O( N/f)$. Second, the behaviour of $\alpha$ for small and large eccentricity is given by
\begin{eqnarray}
\alpha &\sim& -\frac{2 \sqrt{1+\invb^2}}{\invb} \left( \log \psi +1 - 2 \log 2 - \frac{1}{2}\log (1+\mu^2 ) \right)\ \ \textrm{as}\ \ \psi \to 0, \lab{psismall} \\
\alpha &\sim& \frac{(1+\invb^2) \pi}{2 \invb \psi} \ \ \textrm{as}\ \ \psi \to \infty. \lab{psilarge}
\end{eqnarray}
The large-$\psi$ expression \eqn{psilarge} can actually be used to estimate $\alpha$ for values of $\psi$ as small as $1$, which makes it very useful. (For $\mu=1$, for instance, the error in \eqn{psilarge} is 15\%, 10\% and 5\% for $\psi=1, \, 1.5$ and $2$, respectively.) This expression shows in particular that the largest growth rates are attained precisely for $\mu \sim 1$ when $\psi$ is large.
Third, the obvious fact that $\beta > 0$ shows that anticyclonic flows ($\sig = 1$) are more unstable than cyclonic flows ($\sig = -1$). According to \eqn{S}, the growth rate in an anticyclonic flow is a factor $\exp(2 \beta)$ larger than the growth rate of the corresponding cyclonic flow. Formally, this is an $O(1)$ factor, but the typical values of $\beta$ are such that it is numerically very small, so that the instability of cyclones is exceedingly weak and probably negligible in most circumstances. Note that because $\beta$ is a decreasing function of $e$, the asymmetry between cyclones and anticylones is the largest for small eccentricity.  

The formulas \eqn{sigma}--\eqn{beta} give completely explicitly expressions for
the maximum growth rates of the elliptical instability in terms of the three parameters $\eps$, $\invb$ and $e$ (recall that $\psi=\sqrt{e^2-1}$). These growth rates are achieved when the three parameters are related in such a way that $\exp(\i \theta / \eps) = \pm 1$, that is,
\beq \lab{resonance}
\theta = n  \pi \eps, \quad n=1,2,\cdots.
\eeq
This condition can be recognised as a resonance condition between the phase of the inertia-gravity oscillations and
the period of rotation around the elliptical vortex ($2 \pi$ in the dimensionless time used here).

The growth rates can be written more directly in terms of $\eps$, $\invb$  and $e$ by solving \eqn{resonance} peturbatively, with $\theta = \theta_0 + \eps \theta_1 + \cdots$, and $\theta_0$ and $\theta_1$ obtained from \eqn{theta0} and \eqn{theta1}. This gives the approximate position of the instability bands
as well as their width. To leading order, the instability bands are centred at values of $e$ and $\mu$ satisfying
\beq \lab{theta0ex}
\theta_0 = \frac{1}{\mu} \int_{-\pi/2}^{\pi/2} \sqrt{1+\mu^2 + \psi^2 \sin^2 t} \, \d t = \frac{2}{\mu} \int_{0}^{1} \sqrt{\frac{1+\mu^2 + \psi^2 x^2}{1-x^2}} \, \d x = n \pi \eps,  \quad n=1,2,\cdots.
\eeq
The computation of the correction $\eps \theta_1$ is more involved.
Note that it is in principle needed  to obtain a leading-order approximation to the growth rate $\Re \sigma$ as a function of $e$ and $\invb$. This is because the error on $\alpha$ needs to be $o(\eps)$, which requires to approximate the resonance values of $e$ and $\invb$ with an $o(\eps)$ error also.
We do not pursue these detailed computations here: since the values of $e$ and $\mu$ satisfying the resonance condition \eqn{resonance} are $\eps$-close together, \eqn{sigma} provides a useful approximation to the growth rates of the instability without the need to locate the resonances accurately.  This is demonstrated in the next section where we compare the prediction \eqn{sigma} with numerical solutions of the Floquet problem associated with \eqn{zeta1}.

Note that the band width can be deduced directly from the expression \eqn{theta0ex} for $\theta_0$. For fixed $\eps$ and $e$, for instance, $T$ in \eqn{eitheta} can be written as $T = \eps^{-1} \Delta \invb \, \partial_\invb \theta$, where $\Delta \invb$ is the distance between $\invb$ and the resonant values, and the derivative is evaluated at resonance. According to \eqn{lambda}, the instability-band width is therefore $\Delta \mu = 2 \eps |S| / \partial_\mu \theta$ where $\theta$ can be approximated by $\theta_0$.  

\section{Comparison with numerical results}

\begin{figure}[t!]
\begin{center}
\begin{tabular}{cc}
(a) & (b) \\
\includegraphics[width=0.48\textwidth]{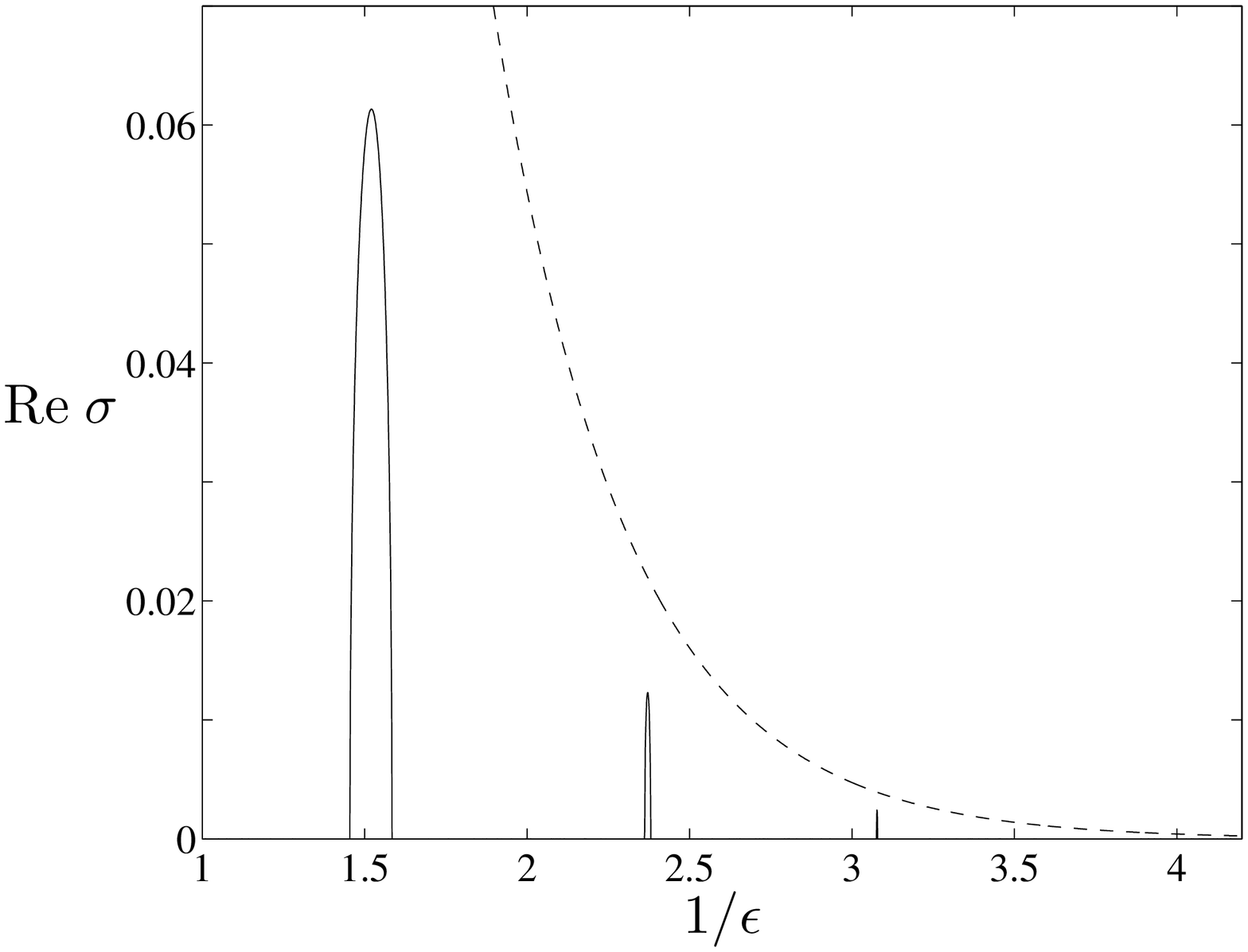} &
\includegraphics[width=0.48\textwidth]{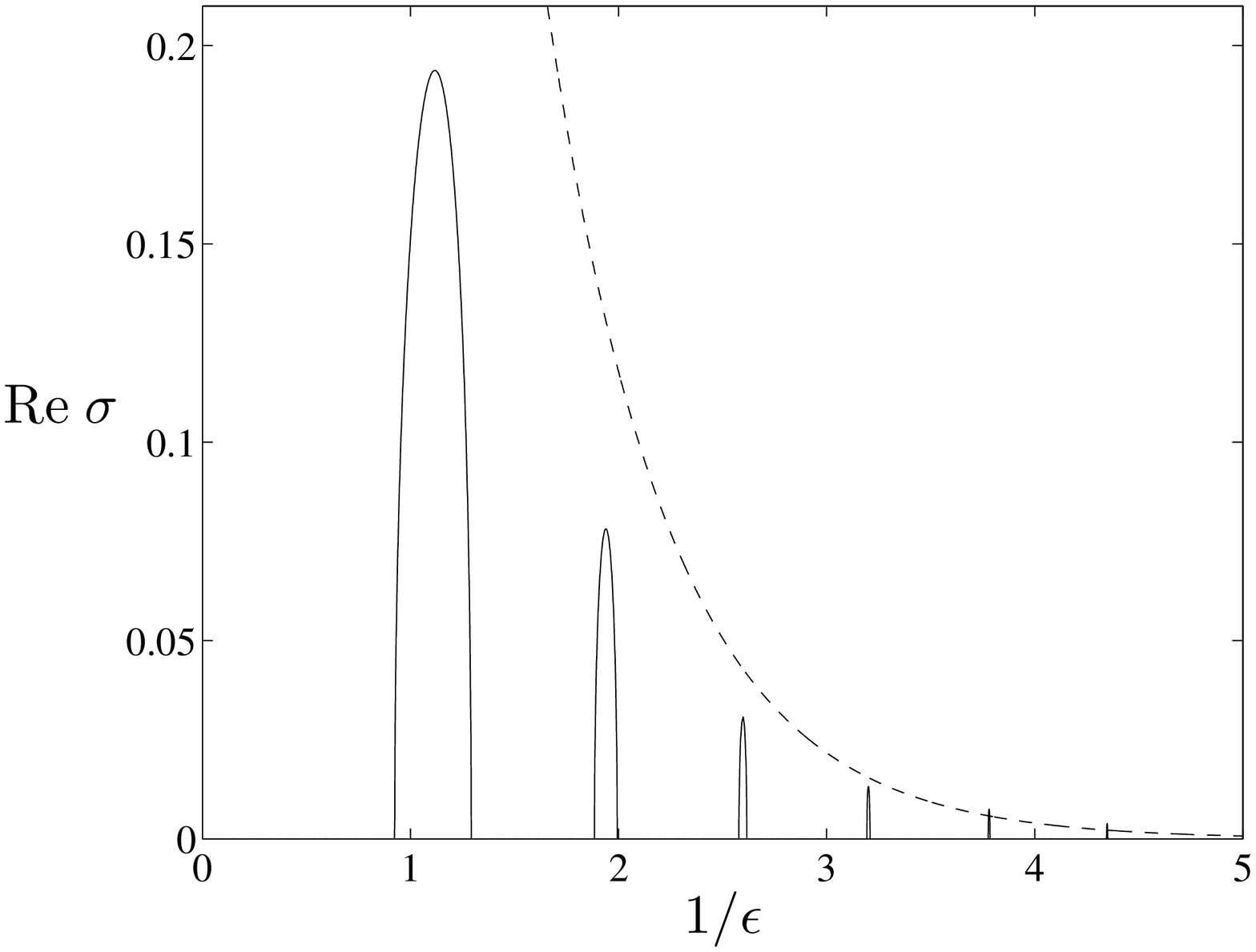}  \smallskip \\
(c) & (d) \\
\includegraphics[width=0.48\textwidth]{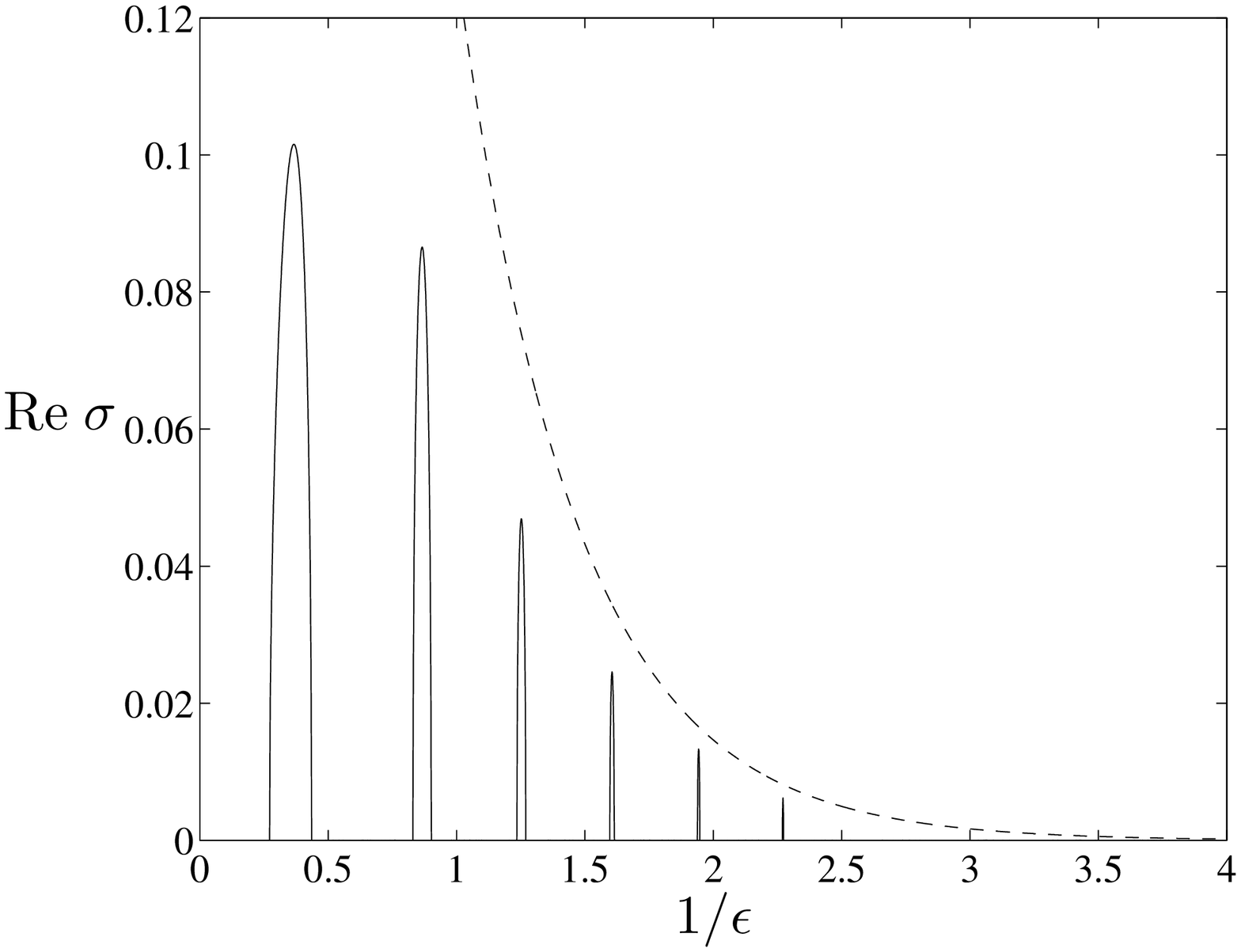} &
\includegraphics[width=0.48\textwidth]{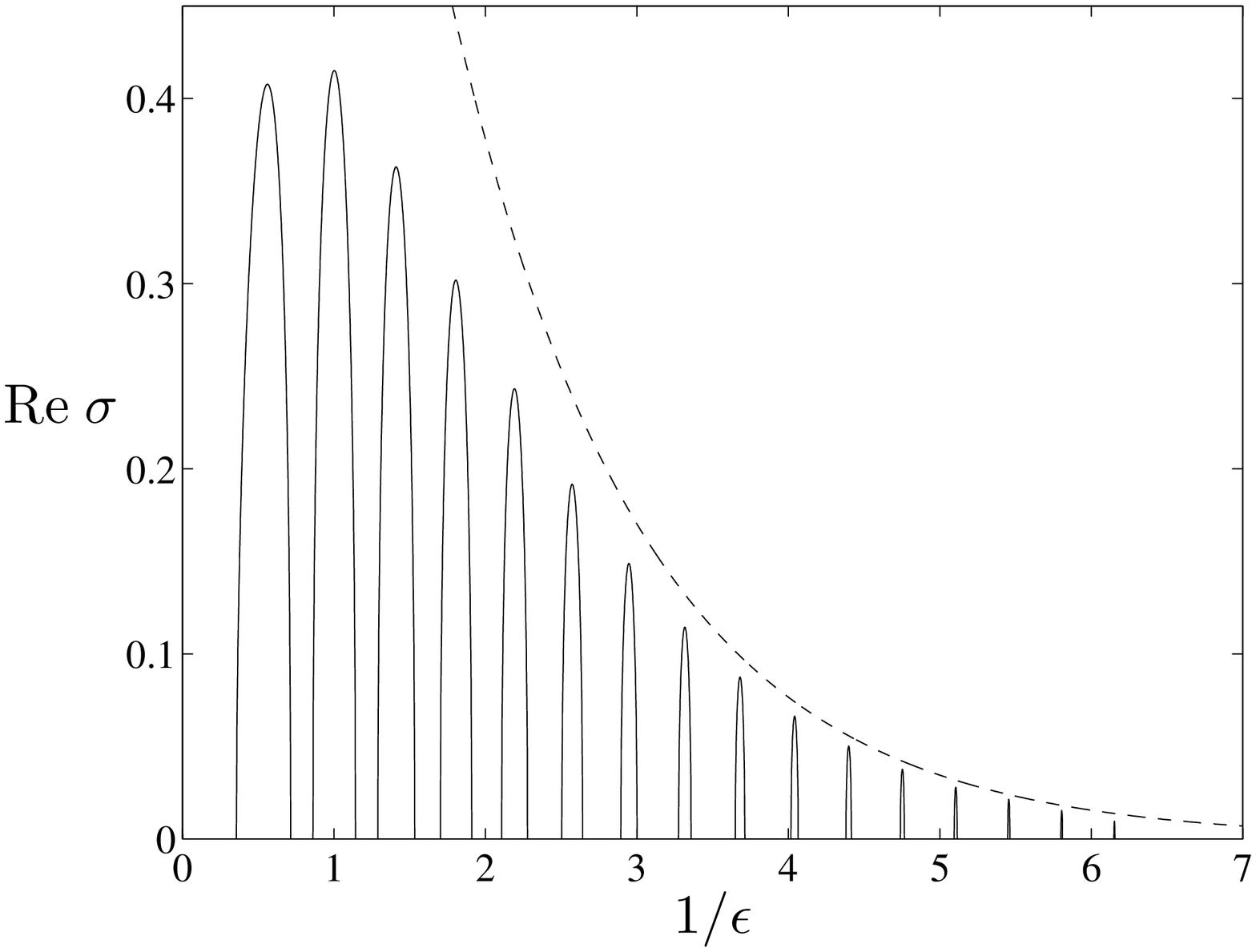}
\end{tabular}
\caption{Growth rates $\Re \sigma$ in anticyclonic flows as functions of the inverse Rossby number $1/\eps$ for (a) $e=1.5, \, \mu=1$; (b) $e=2, \, \mu=1$; (c) $e=2, \, \mu=0.5$; and (d) $e=4, \, \mu=1$. The growth rates computed numerically (solid lines) are compared with the asymptotic estimate of the maximum growth rates $\sigma_\mathrm{max}$ (dashed lines).}
\label{anticyclonic_rates}
\end{center}
\end{figure}

\begin{figure}[t]
\begin{center}
\begin{tabular}{cc}
(a) & (b) \\ 
\includegraphics[width=0.48\textwidth]{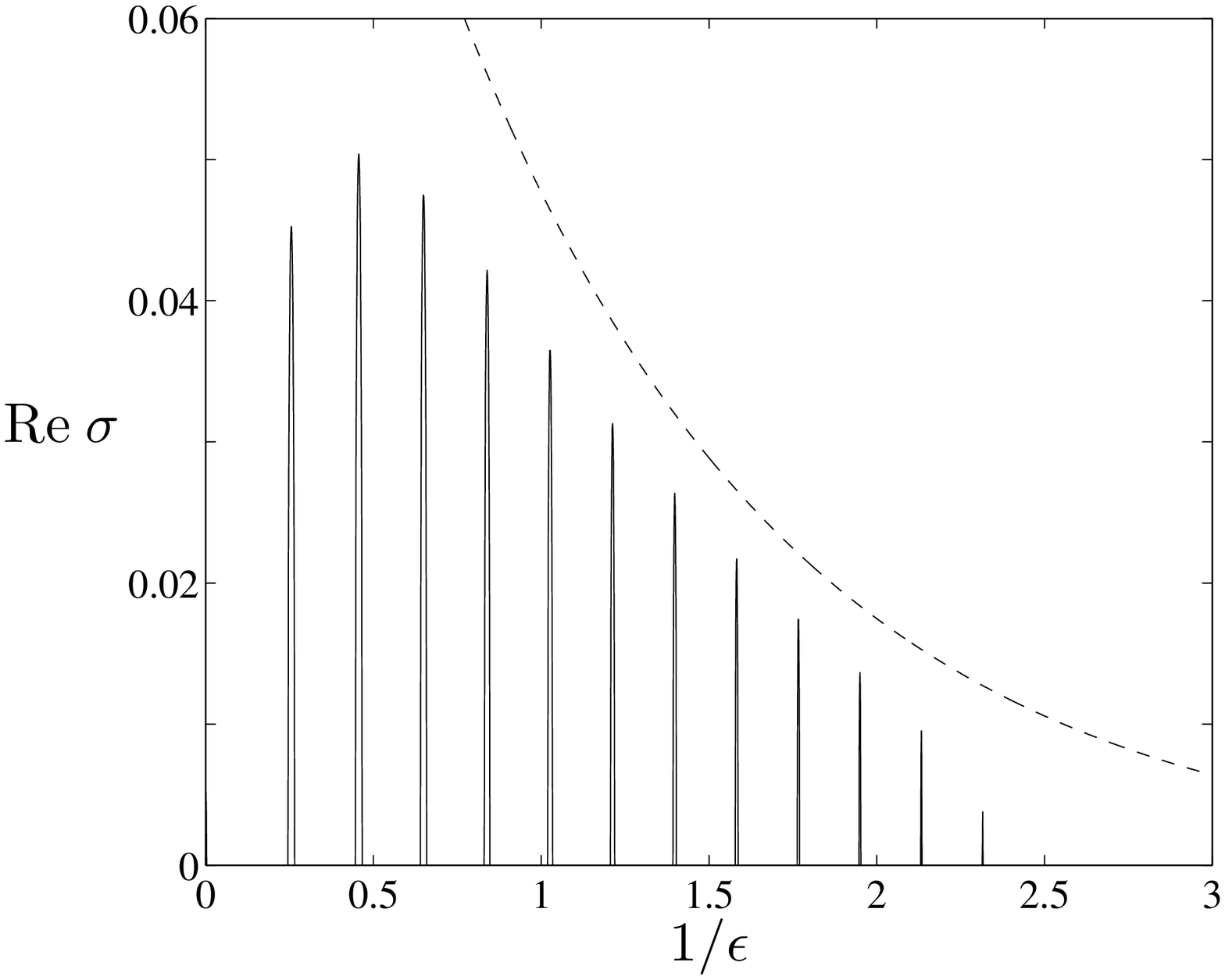}
&
\includegraphics[width=0.48\textwidth]{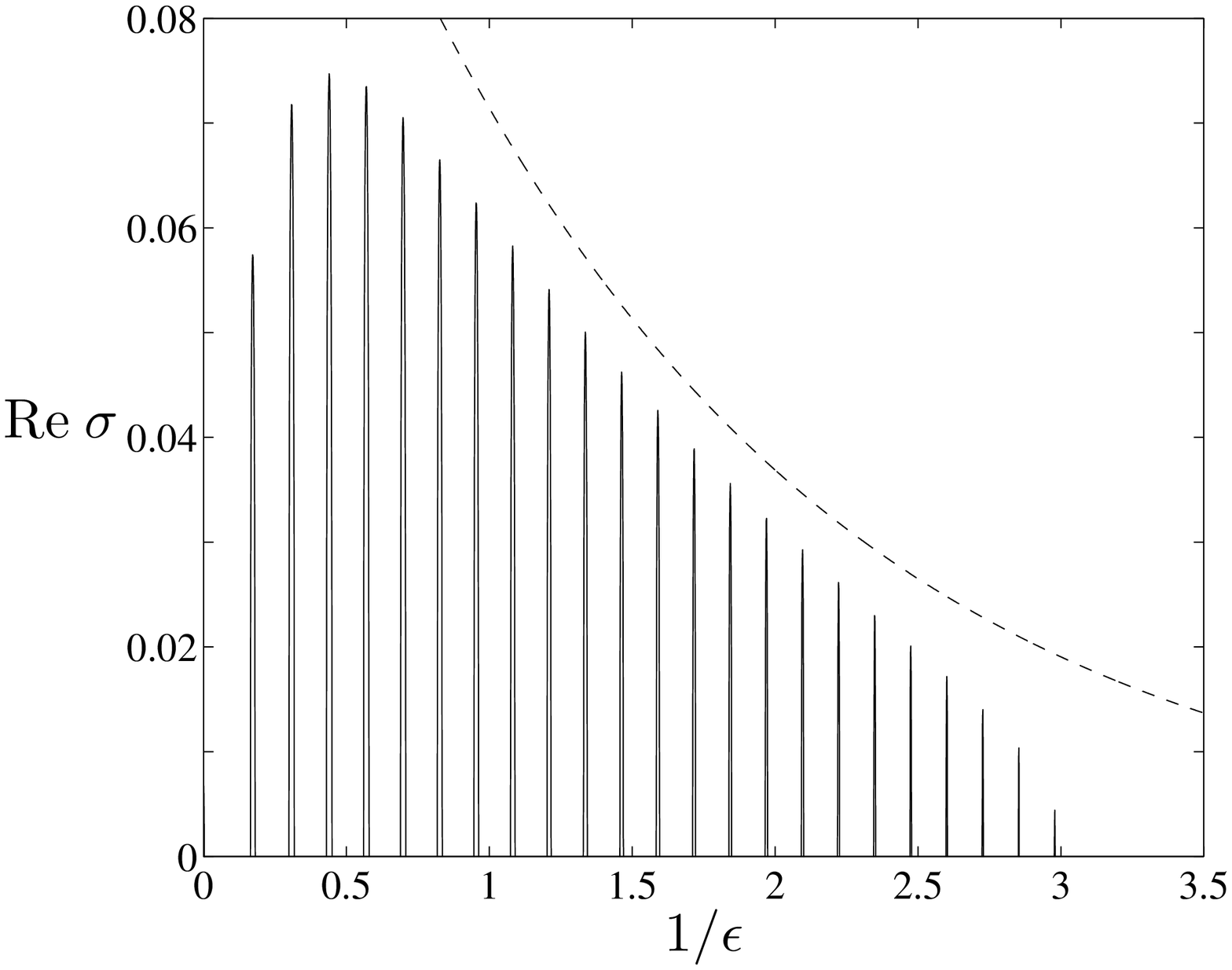}
\end{tabular}
\caption{Growth rates $\Re \sigma$ in cyclonic flows as functions of the inverse Rossby number $1/\eps$ for (a) $e=4, \, \mu=0.5$; and (b) $e=6, \, \mu=0.5$. The growth rates computed numerically (solid lines) are compared with the asymptotic estimate of the maximum growth rates $\sigma_\mathrm{max}$ (dashed lines).}
\label{cyclonic_rates}
\end{center}
\end{figure}

We have solved the Floquet problem associated with equation \eqn{zeta2} for the amplitude $\zeta$ numerically using Matlab's standard Runge--Kutta solver. The growth rates $\Re \sigma$ obtained in this manner are compared with the asymptotic 
estimate \eqn{sigma} for $\sigma_\mathrm{max}$. To emphasise the exponential dependence on the inverse Rossby number $1/\eps$, it is convenient to display $\Re \sigma$ as a function of $1/\eps$ for fixed values of $\mu$ and of $e$. 
Figure \ref{anticyclonic_rates} shows the results obtained in case of anticyclonic vortices (with $\varsigma = 1$) for several values of $e$ and $\mu$.  Similar results for cyclonic vortices ($\varsigma = -1$) are displayed in Figure \ref{cyclonic_rates}. 

The figures confirm the validity of our asymptotic estimate. They also suggest that this estimate remains useful for moderately small values of $\eps$, say $\eps \lesssim 1/2$.
Note that the dimensional growth rates are obtained by multiplying $\sigma$ by $\sqrt{ab}$ which is related to the relative vorticity $\Omega=a+b$ of the flow by $\sqrt{ab}=\Omega/(e+e^{-1})$. As expected from our asymptotics, the growth rates in the case of cyclonic flows are exceedingly small for $\eps \ll 1$ even for the large eccentricities used in Figure \ref{cyclonic_rates}.  Nonetheless, our results clarify the fact that all elliptical flows are unstable, regardless of the sense of the rotation, of its strength, and of the strength of the stratification.  Note that the match between asymptotic and numerical results for cyclonic flows appears to degrade for small $\eps$ (i.e., large $1/\eps$); this is because the  smallness of both the growth rate and instability-band width makes the maximum growth rate delicate to estimate numerically.  

The separation between instability bands can be estimated from the asymptotic formula \eqn{theta0ex}: in terms of the varying $1/\eps$ used in the figures, it is given by
\[
\gamma = \frac{\pi \mu}{2 \displaystyle{\int_{0}^{1} \sqrt{\frac{1+\mu^2 + \psi^2 x^2}{1-x^2}} \, \d x}}.
\]
Evaluating this quantity for the parameters chosen for the figures gives $\gamma=0.62, \, 0.54,\, 0.31$ and $0.34$ for the parameters of Figure \ref{anticyclonic_rates} (a)--(d), and $\gamma=0.18$ and $0.12$ for the parameters of Figure \ref{cyclonic_rates} (a) and (b), in good agreement with the numerical results. 

\begin{figure}[t]
\begin{center}
\includegraphics[width=0.48\textwidth]{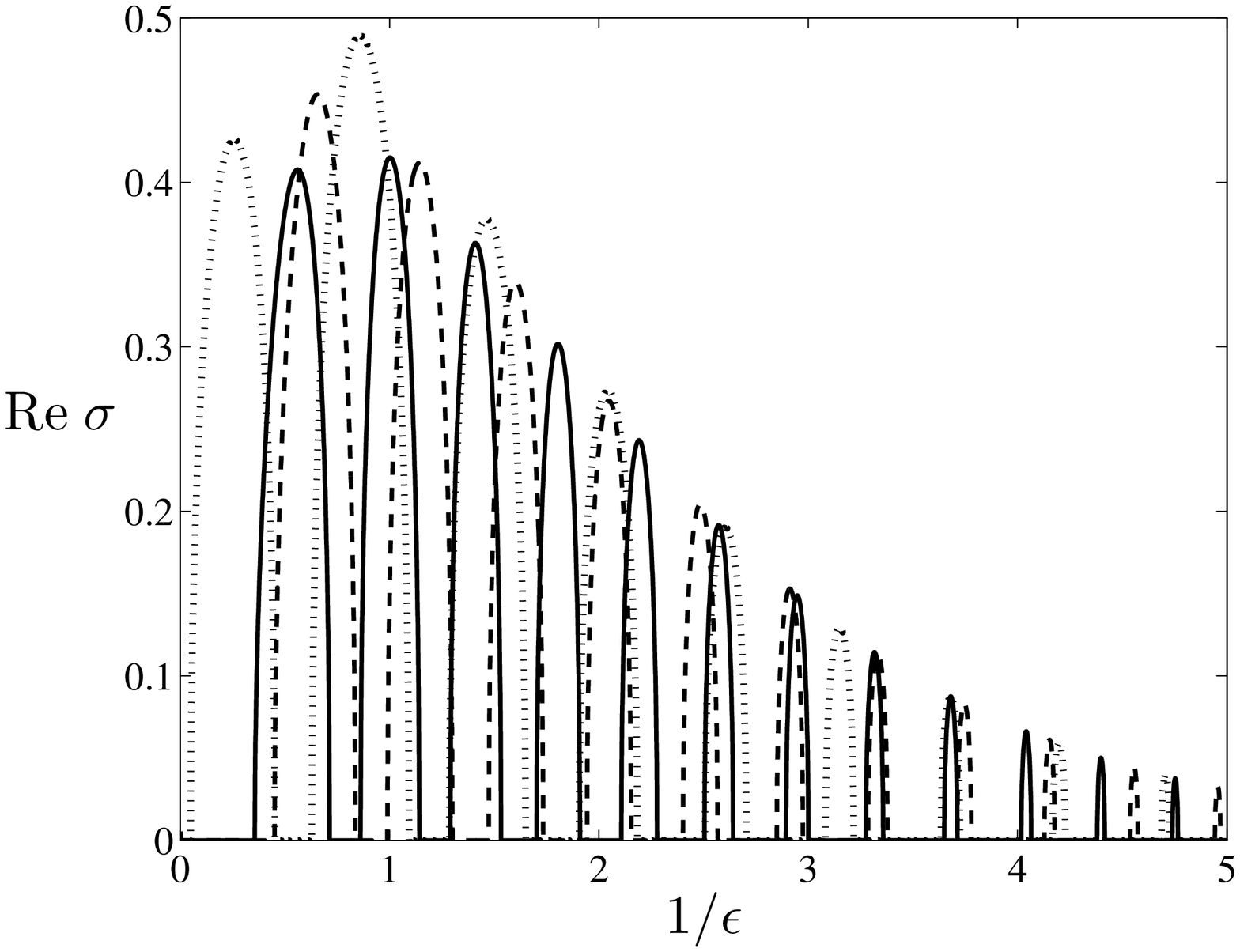}
\caption{Effect of the hydrostatic approximation: the growth rate $\Re \sigma$ is plotted as a function of the inverse Rossby number $1/\eps$ for an anticyclonic flow with $e=4$ and $\mu=1$, in the hydrostatic limit $N/f \to \infty$ (solid lines), for $N/f=6$ (dashed lines), and for $N/f=3$ (dotted lines).}
\label{comparison}
\end{center}
\end{figure}

Our derivation of an asymptotic expression for the growth rate makes the hydrostatic approximation, which assumes that $N \gg f$, $m_0 \gg k_0$ and $\mu=f m_0/(N k_0) =O(1)$. This assumption, which could be relaxed, is made because it corresponds to the regime most relevant to atmospheric and oceanic flows; it is consistent in the sense that the growth rates obtained are maximized for $\mu=O(1)$ and decay rapidly for $\mu \gg 1$ or $\mu \ll 1$. To test the sensitivity of the results to the hydrostatic approximation, we have solved the Floquet problem numerically for the full, unapproximated equation \eqn{nonhydro} 
for moderately large values of $N/f$ and $m_0/k_0$. The results obtained for $\mu=1$ and $e=4$ are displayed in  Figure \ref{comparison}. This compares the growth rate obtained in the hydrostatic approximation with those obtained for $N/f=m_0/k_0=3$ and $6$. Except for $\eps \gtrsim 1$, there is relatively little difference between the results: the maximum growth rates fall on the same curve, well described by the hydrostatic asymptotics. Of course, the location of the instability bands changes depending on $S$, but this is not significant, since they would also change if $m_0/k_0$ was varied independently of $N/f$, as is physically relevant. 
\medskip

\noindent
\textbf{Acknowledgments.} The authors acknowledge the support of the EPSRC Network `Wave--Flow Interactions'. JMA is supported by a studentship of the UK Natural Environment Research Council.

\appendix
\section{Exponential asymptotics} \label{sec:expasy}

In this Appendix, we compute the (exponentially small) Stokes multiplier $S$ which quantifies the switching on of one branch of the WKB solution by the other (see \eqn{wkb-}--\eqn{wkb+}) through a Stokes phenomenon\citep{ablo-foka,pari-wood}.
The Stokes phenomenon  is associated with the presence of complex turning points, that is, complex times where $\omega = 0$. From \eqn{omega}, these are located at
\[
t_n = \i \sinh^{-1} \frac{\sqrt{1+\invb^2}}{\psi} + n \pi, \quad n=0,\pm1,\pm2,\cdots
\]
and $\bar t_n$. The Stokes phenomenon occurs as $t \in \mathbb{R}$ crosses one of the Stokes lines joining $t_n$ to $\bar t_n$. In the interval $[-\pi/2,\pi/2]$ of interest, the  only Stokes line is the segment of $\Re t=0$ joining $t_0$ to $\bar t_0$. We compute $S$ using matched asymptotics, examining how the solution $\zeta = A(t) \exp(\i \theta(t)/\eps)$ connects to the solution $\zeta = A(t) [\exp(\i \theta(t)/\eps) + S \exp(-\i \theta(t)/\eps)]$ as this segment is crossed.\citep{haki98}

To analyse the behaviour of the solution in the neighbourhood of $t_0$, we first note that
\beq \lab{omt0}
\omega \sim a^{1/2} \e^{\i \pi/4} (t-t_0)^{1/2}, \inter{where} a=\sqrt{2(1+\invb^{2})(1+\psi^2+\invb^{2})}/\invb^2,
\eeq
as $t \to t_0$.
We then rescale the evolution equation for $\zeta$ near $t_0$ by defining the inner variables
\[
\tau = \eps^{-2/3} a^{1/3} (t - t_0) \inter{and} Z(\tau)=\zeta(t).
\]
Retaining only the leading order terms, this transforms \eqn{zeta2} into
the equation
\beq \lab{Z}
\dt{^2 Z}{\tau^2} + \i \tau Z = 0.
\eeq
Solutions can be written in terms of the Airy functions $\Ai(\e^{-\i \pi/6} \tau)$ and $\Bi(\e^{-\i \pi/6} \tau)$. We claim that the solution of interest is
\beq \lab{Z1}
Z \sim C \left[\Ai(\e^{-\i \pi/6} \tau) + \i \Bi(\e^{-\i \pi/6} \tau)\right].
\eeq
We verify that this solution matches  $A(t) \exp(\i \theta(t)/\eps)$ to the left of the Stokes line $\Re t =0$; in doing so we find an expression for the constant $C$. It is convenient to verify the matching along the line $\ph \tau = -5 \pi/6$; this is an anti-Stokes line along which the two independent solutions of \eqn{Z} have similar orders of magnitudes. Along this line, we can use the asymptotic formulas\citep{abra-steg}
\begin{eqnarray}
\Ai(-z) &\sim& \frac{1}{\sqrt{\pi} z^{1/4}} \cos(2 z^{3/2}/3-\pi/4), \lab{ai} \\
\Bi(-z) &\sim& -\frac{1}{\sqrt{\pi} z^{1/4}} \sin(2 z^{3/2}/3-\pi/4),
\end{eqnarray}
with $z = - \exp(-\i \pi/6) \tau = \exp(5 \i \pi/6) \tau \to +\infty$.
Thus we have
\beq \lab{Zin}
Z \sim \frac{C \e^{\i \pi/24}}{\sqrt{\pi} \tau^{1/4}} \e^{2\i \e^{\i\pi/4} \tau^{3/2}/3}.
\eeq
On the other hand, using \eqn{a}--\eqn{theta1} the solution $\zeta = A(t) \exp(\i \theta(t))$, valid in the outer region away from $t_0$ and to the left of the Stokes line  $\Re t =0$, can be written as
\begin{eqnarray}
\zeta &\sim& \frac{1}{\omega^{1/2}} \e^{\int_{-\pi/2}^ t p(t') \, \d t'/2} \e^{\i \eps^{-1} \int_{-\pi/2}^t \left[\omega(t')- \eps q(t')/(2 \omega(t'))\right] \, \d t'} \nonumber \\
&\sim& \frac{\e^{-\i \pi/8}}{(\eps a)^{1/6} \tau^{1/4}}
\e^{\int_{\Gamma_-} p(t') \, \d t'/2}
 \e^{\i \eps^{-1} \int_{\Gamma_-}  \left[\omega(t')- \eps q(t')/(2 \omega(t'))\right] \, \d t'} \e^{2\i \e^{\i\pi/4} \tau^{2/3}/3}, \lab{zetaout-}
\end{eqnarray}
after using \eqn{omt0}.
Here $\Gamma_-$ denotes a contour joining $-\pi/2$ to $t_0$ while remaining to the left of the Stokes line $\Re t =0$.
Comparing \eqn{zetaout-} with \eqn{Zin} shows that $\zeta$ correctly matches $Z$ provided that
\beq \lab{C}
C = \frac{\sqrt{\pi} \e^{-\i \pi/6}}{(\eps a)^{1/6}} \e^{\int_{\Gamma_-} p(t') \, \d t'/2}
 \e^{\i \eps^{-1} \int_{\Gamma_-}  \left[\omega(t')- \eps q(t')/(2 \omega(t'))\right] \, \d t'}.
 \eeq

We now match $Z$ with the outer solution valid to  the right of the Stokes line $\Re t=0$.
A connection formula for Airy functions\citep{abra-steg} gives the alternative form
\beq \lab{Z2}
Z \sim 2 C \e^{\i \pi/3} \Ai(\e^{-\i 5 \pi/6} \tau),
\eeq
for \eqn{Z1}.
Carrying out the matching on the anti-Stokes line $\ph \tau = \exp(-\i \pi/6)$, we can use the asymptotic formula for $\Ai$ in \eqn{ai} to write that
\beq \lab{Zin2}
Z \sim  \frac{C \e^{\i \pi/24}}{\sqrt{\pi}\tau^{1/4}} \left(\e^{2 \i \e^{\i \pi/4} \tau^{3/2}/3} + \i\e^{-2 \i \e^{\i \pi/4} \tau^{3/2}/3}\right)
\eeq
for $|\tau| \to \infty$.
This should be matched with the form $\zeta(t)=A(t)[\exp(\i \theta(t)/\eps)+S \exp(-\i \theta(t)/\eps)]$ of the solution to the right of the Stokes line. Using \eqn{omt0} gives the asymptotics
\begin{eqnarray*}
\zeta &\sim& \frac{1}{\omega^{1/2}} \e^{\int_{-\pi/2}^ t p(t') \, \d t'/2} \left[ \e^{\i \eps^{-1} \int_{-\pi/2}^t \left[\omega(t')- \eps q(t')/(2 \omega(t'))\right] \, \d t'}
+  S \e^{-\i \eps^{-1} \int_{-\pi/2}^t \left[\omega(t')- \eps q(t')/(2 \omega(t')) \right] \, \d t'} \right] \\
&\sim& \frac{\e^{-\i \pi/8}}{(\eps a)^{1/6} \tau^{1/4}}
\e^{\int_{\Gamma_+} p(t') \, \d t'/2}
\left[
 \e^{\i \eps^{-1} \int_{\Gamma_+}  \left[\omega(t')- \eps q(t')/(2 \omega(t'))\right] \, \d t'} \e^{2\i \e^{\i\pi/4} \tau^{3/2}/3} \right. \\
 && \qquad \qquad \qquad \qquad \qquad \qquad  \left. + S  \e^{-\i \eps^{-1} \int_{\Gamma_+}  \left[\omega(t')- \eps q(t')/(2 \omega(t'))\right] \, \d t'} \e^{-2\i \e^{\i\pi/4} \tau^{3/2}/3} \right],
\end{eqnarray*}
where $\Gamma_+$ denotes a contour joining $-\pi/2$ to $t_0$. This contour crosses the Stokes line $\Re t =0$ below the singular point $t_\mathrm{p}$ of $p(t)$ and $q(t)$, given by $t_\mathrm{p}=\i \sinh^{-1}(1/\psi)$.
Taking \eqn{C} into account, the matching with \eqn{Zin2} gives
the two equations
\begin{eqnarray}
\e^{\int_{\Gamma_-} p(t) \, \d t/2} \e^{-\i \int_{\Gamma_-} q(t)/\omega(t) \, \d t /2} &=& \e^{\int_{\Gamma_+} p(t) \, \d t/2} \e^{-\i \int_{\Gamma_+} q(t)/\omega(t) \, \d t/2}, \lab{1} \\
\i \e^{\int_{\Gamma_-} p(t') \, \d t'/2} \e^{\i \eps^{-1} \int_{\Gamma_-}  \left[\omega(t')- \eps q(t')/(2 \omega(t'))\right] \, \d t'}&=& S \e^{\int_{\Gamma_+} p(t') \, \d t'/2} \e^{-\i \eps^{-1} \int_{\Gamma_+}  \left[\omega(t')- \eps q(t')/(2 \omega(t'))\right] \, \d t'}.  \lab{2}
\end{eqnarray}
We can now deform the integration contours. The difference $\int_{\Gamma_+}-\int_{\Gamma_-}$ reduces to an integral along a closed contour encircling $t_\mathrm{p}$. Computing the corresponding residues using \eqn{fg} gives
\[
\Res_{t_\mathrm{p}} p = 1 \inter{and} \Res_{t_\mathrm{p}} \frac{q}{\omega} = - \i \sig,
\]
Taking this into account confirms that \eqn{1} is an identity. Using $\Res_{t_\mathrm{p}} p = 1$ in \eqn{2} gives the Stokes multiplier as
\[
S = - \i  \e^{2 \i \eps^{-1} \dashint_{-\pi/2}^{t_0} \left[ \omega(t')- \eps q(t')/(2 \omega(t'))\right] \, \d t'},
\]
where the Cauchy principal value integral, denoted by $\dashint$, is necessary because $q(t)$ has a pole at $t=t_\mathrm{p}$.
It follows that $|S|$, giving the instability growth rate, can be written as
\[
|S|= \e^{-\alpha/\eps + \sig \beta},
\]
where the two constants
\[
\alpha = - 2 \i \int_0^{t_0} \omega(t) \, \d t \inter{and}
\beta = -\i \dashint_0^{t_0} \frac{ \sig q(t)}{\omega(t)} \ \d t
\]
are real, positive and independent of $\sig$. Using \eqn{omega} and \eqn{fg}, they can be given the more explicit forms
\[
\alpha = \frac{2}{\invb} \int_0^{\sinh^{-1}(\sqrt{1+\invb^{2}}/\psi)} \sqrt{1- \psi^2 \sinh^2 u + \invb^2} \, \d u,
\]
and
\[
\beta = \invb \sig \dashint_0^{\sinh^{-1}(\sqrt{1+\invb^2}/\psi)} \left( e + e^{-1} + \frac{2 e}{1-\psi^2 \sinh^2 u} \right) \frac{\d u}{\sqrt{1- \psi^2 \sinh^2 u+\invb^2}},
\]
and further transformed into the convenient expressions \eqn{alpha}--\eqn{beta}.

\bibliographystyle{unsrtnat}
\bibliography{mybib}

\begin{thebibliography}{18}
\providecommand{\natexlab}[1]{#1}
\providecommand{\url}[1]{\texttt{#1}}
\expandafter\ifx\csname urlstyle\endcsname\relax
  \providecommand{\doi}[1]{doi: #1}\else
  \providecommand{\doi}{doi: \begingroup \urlstyle{rm}\Url}\fi

\bibitem[Kerswell(2002)]{kers02}
R.~R. Kerswell.
\newblock Elliptical instability.
\newblock \emph{Ann. Rev. Fluid Mech.}, 34:\penalty0 83--113, 2002.

\bibitem[Miyazaki(19993)]{miya93}
T.~Miyazaki.
\newblock Elliptical instability in a stably stratified rotating fluid.
\newblock \emph{Phys. Fluids}, A 5:\penalty0 2702--2709, 19993.

\bibitem[McWilliams and Yavneh(1998)]{mcwi-yavn}
J.~C. McWilliams and I.~Yavneh.
\newblock Fluctuation growth and instability associated with a singularity of
  the balance equations.
\newblock \emph{Phys. Fluids}, 10:\penalty0 2587--2596, 1998.

\bibitem[Molemaker et~al.(2005)Molemaker, McWilliams, and Yavneh]{mole-et-al05}
M.~J. Molemaker, J.~C. McWilliams, and I.~Yavneh.
\newblock Baroclinic instability and loss of balance.
\newblock \emph{J. Phys. Oceanogr.}, 35:\penalty0 1505--1517, 2005.

\bibitem[Vanneste and Yavneh(2007)]{v-yavn07}
J.~Vanneste and I.~Yavneh.
\newblock Unbalanced instabilities of rapidly rotating stratified shear flows.
\newblock \emph{J. Fluid Mech.}, 584:\penalty0 373--396, 2007.

\bibitem[McWilliams et~al.(2004)McWilliams, Molemaker, and
  Yavneh]{mcwi-et-al04}
J.~C. McWilliams, M.~J. Molemaker, and I.~Yavneh.
\newblock Ageostrophic, anticyclonic instability of a geostrophic, barotropic
  boundary current.
\newblock \emph{Phys. Fluids}, 16:\penalty0 3720--3725, 2004.

\bibitem[Plougonven et~al.(2005)Plougonven, Muraki, and Snyder]{plou-et-al}
R.~Plougonven, D.~J. Muraki, and C.~Snyder.
\newblock A baroclinic instability that couples balanced motions and gravity
  waves.
\newblock \emph{J. Atmos. Sci.}, 62:\penalty0 1545--1559, 2005.

\bibitem[Vanneste and Yavneh(2004)]{v-yavn04}
J.~Vanneste and I.~Yavneh.
\newblock Exponentially small inertia-gravity waves and the breakdown of
  quasi-geostrophic balance.
\newblock \emph{J. Atmos. Sci.}, 61:\penalty0 211--223, 2004.

\bibitem[Vanneste(2008)]{v08}
J.~Vanneste.
\newblock Exponential smallness of inertia-gravity-wave generation at small
  {R}ossby number.
\newblock \emph{J. Atmos. Sci.}, 65:\penalty0 1622--Ð1637, 2008.

\bibitem[Bender and Orszag(1999)]{bend-orsz}
C.~M. Bender and S.~A. Orszag.
\newblock \emph{Advanced mathematical methods for scientists and engineers}.
\newblock Springer, 1999.

\bibitem[Weinstein and Keller(1987)]{wein-kell87}
M.~I. Weinstein and J.~B. Keller.
\newblock Asymptotic behaviour of stability regions for {H}ill's equation.
\newblock \emph{SIAM J. Appl. Math.}, 47:\penalty0 941--958, 1987.

\bibitem[Ablowitz and Fokas(1997)]{ablo-foka}
M.~J. Ablowitz and A.~S. Fokas.
\newblock \emph{Complex variables: introduction and applications}.
\newblock Cambridge University Press, 1997.

\bibitem[Friedlander and Lipton-Lifschitz(2003)]{frie-lipt}
S.~J. Friedlander and A.~Lipton-Lifschitz.
\newblock Localized instabilities in fluids.
\newblock In S.J. Friedlander and D.~Serre, editors, \emph{Handbook of
  Mathematical Fluid Dynamics, vol. II}, pages 289--353. Elsevier Science,
  2003.

\bibitem[{Le Diz{\`e}s}(2000)]{ledi00}
S.~{Le Diz{\`e}s}.
\newblock Three-dimensional instability of a multipolar in a rotating flow.
\newblock \emph{Phys. Fluids}, 12:\penalty0 2762--2774, 2000.

\bibitem[Kloosterziel et~al.(2007)Kloosterziel, Carnevale, and
  Orlandi]{kloo-et-al}
R.~C. Kloosterziel, G.~F. Carnevale, and P.~Orlandi.
\newblock Inertial instability in rotating stratified fluids: barotropic
  vortices.
\newblock \emph{J. Fluid Mech.}, 583:\penalty0 379--412, 2007.

\bibitem[Paris and Wood(1995)]{pari-wood}
R.~B. Paris and A.~D. Wood.
\newblock Stokes phenomenon demystified.
\newblock \emph{Bull. Inst. Math. Appl.}, 31:\penalty0 21--28, 1995.

\bibitem[Hakim(1998)]{haki98}
V.~Hakim.
\newblock Asymptotic techniques in nonlinear problems: some illustrative
  examples.
\newblock In C.~Godr\`{e}che and P.~Manneville, editors, \emph{Hydrodynamics
  and nonlinear instabilities}, chapter~3, pages 295--386. Cambridge University
  Press, 1998.

\bibitem[Abramowitz and Stegun(1965)]{abra-steg}
M.~Abramowitz and I.~A. Stegun.
\newblock \emph{Handbook of mathematical functions}.
\newblock Dover, 1965.

\end{thebibliography}

\end{document}